{%% 
\documentclass[final,5p
,twocolumn,authoryear]{elsarticle}

\usepackage{amsmath,amsfonts,amssymb,mathtools}
\usepackage{hyperref}% add hypertext capabilities
\usepackage[nameinlink,capitalise]{cleveref}
\usepackage{bm}
\usepackage{xcolor}

\usepackage[scr]{rsfso}
\newcommand{\pd}{\partial}
\newcommand{\de}{\mathrm{d}}
\newcommand{\cH}{\mathcal{H}}
\newcommand{\cZ}{\mathcal{Z}}
\renewcommand{\pd}{\partial}
\newcommand{\ii}{\mathrm{i}}
\newcommand{\bmag}{\mathcal{Q}}
\newcommand{\bevo}{\mathcal{E}}

\newcommand{\physrep}{Physics Reports}
\newcommand{\jcap}{J. Cosmol. Astrop. Phys.}
\newcommand{\mnras}{Mon. Not. Roy. Astron. Soc.}
\newcommand{\aap}{Astron. Astrophys.}
\newcommand{\prd}{Physical Review D}
\newcommand{\prl}{Physical Review Letters}
\newcommand{\aj}{AJ}
\newcommand{\apj}{ApJ}

\newcommand{\fluxum}{\times10^{-16} \rm \; erg\,cm^{-2}\,s^{-1}}

\def\review#1{\textcolor{black}{#1}}
\usepackage{cancel}

\journal{Physics of the Dark Universe}

\begin{document}

\begin{frontmatter}



\title{Detecting Relativistic Doppler in Galaxy Clustering with Tailored Galaxy Samples} %% Article title


\author[label1,label2]{Federico Montano}
\ead{federico.montano@unito.it}
\author[label1,label2,label3,label4]{Stefano Camera}
\ead{stefano.camera@unito.it}

\affiliation[label1]{organization={Dipartimento di Fisica, Università degli Studi di Torino},
             addressline={Via P.\ Giuria 1},
             city={Torino},
             postcode={10125},
             country={Italy}}
\affiliation[label2]{organization={INFN -- Istituto Nazionale di Fisica Nucleare, Sezione di Torino},
             addressline={Via P.\ Giuria 1},
             city={Torino},
             postcode={10125},
             country={Italy}}
\affiliation[label3]{organization={INAF -- Istituto Nazionale di Astrofisica, Osservatorio Astrofisico di Torino},
             addressline={Strada Osservatorio 20},
             city={Pino Torinese},
             postcode={10025},
             country={Italy}}
\affiliation[label4]{organization={Department of Physics \& Astronomy, University of the Western Cape},
             city={Cape Town},
             postcode={7535},
             country={South Africa}}

%% Abstract
\begin{abstract}
We present a method to obtain a high-significance detection of relativistic effects on cosmological scales. Measurements of such effects would be instrumental for our understanding of the Universe, as they would provide a further confirmation of the validity of general relativity as the correct description of the gravitational interaction, in a regime very far from that of strong gravity, where it has been tested to exquisite accuracy. Despite its relevance, the detection of relativistic effects has hitherto eluded us, mainly because they are stronger on the largest cosmic scales, plagued by cosmic variance. Our work focuses on the cosmological probe of galaxy clustering, describing the excess probability of finding pairs of galaxies at a given separation due to them being part of the same underlying cosmic large-scale structure. We focus on the two-point correlation function of the distribution of galaxies in Fourier space---the power spectrum---where relativistic effects appear as an imaginary contribution to the real power spectrum. By carefully tailoring cuts in magnitude/luminosity, we are able to obtain two samples (bright and faint) of the same galaxy population, whose cross-correlation power spectrum allows for a detection of the relativistic contribution. In particular, we optimise the definition of the samples to maximise the detection significance of the relativistic Doppler term for both a low-$z$ Bright Galaxy Sample and a high-$z$ H\(\alpha\) emission line galaxy population.
\end{abstract}



%% Keywords

\end{frontmatter}



%% Use \subsubsection, \paragraph, \subparagraph commands to 
%% start 3rd, 4th and 5th level sections.
%% Refer following link for more details.
%% https://en.wikibooks.org/wiki/LaTeX/Document_Structure#Sectioning_commands
\section{Introduction}
\label{sec1intro}
Gravity is the force that mostly drives the evolution of the Universe, since it is the fundamental interaction that can act on cosmic distances. Therefore, our understanding of its properties is paramount for cosmology. As such, the current concordance cosmological model is rooted in the theory of general relativity (GR). Despite GR being supported by several stunning experimental observations \citep{2010JHA....41...41H,1968PhRvL..20.1265S,2002PhT....55e..41A,1975ApJ...195L..51H,2005ASPC..328...25W,2023ApJ...951L...8A,2016PhRvL.116f1102A}, it is still poorly tested in the extremely-weak field regime of cosmological scales. In this context, a measurement of an effect due to GR coming from cosmology would represent another amazing success of Einsteinian gravity, whereas departures from GR would be a window into the intriguing `modified gravity' scenario \citep{2012PhR...513....1C,2016ARNPS..66...95J}. 

The large-scale structure of the Universe offers an important test bench for gravity theories. The properties of galaxy clustering tell us about the driving force of cosmological evolution, simply because a clumpy Universe like the one we live in has to be the result of ages of gravitational accretion, starting from nearly-homogeneous primordial stages. In particular, a statistical investigation of the distribution of a certain tracer, e.g.\ a type of galaxy, can display signatures of various effects that affect the observed position and magnitude of sources in the sky. We ascribe most of these phenomena to the fact that galaxy surveys actually map cosmic structures in observed redshift space, rather than in real space. Also, gravitational lensing is known to affect the observed clustering of sources, because photons from distant galaxies are scattered by the intervening large-scale structure on their journey towards us. Other corrections that should be taken into account are so-called relativistic effects, amongst which the dominant one is a relativistic Doppler term. The amplitude of such Doppler term is itself affected by gravitational lensing effects and the evolution of the target galaxy population.

In this paper, we present a strategy to detect the relativistic Doppler through the study of galaxy clustering. Following the idea first proposed by \citet[][see also \citealp{2016JCAP...08..021B, 2017JCAP...01..032G}]{2014PhRvD..89h3535B}, we divide a galaxy population according to observed flux density (namely, according to luminosity) in order to work with two complementary selections of the same sample. The faint and bright samples thus obtained are by construction independent and are hence two promising candidates for a cross-correlation study---i.e.\ the analysis of the covariance between the galaxy distributions in the two sub-samples.

Since relativistic effects are subdominant on small scales, which are also most affected by the non-linear growth of structures, it is convenient to isolate small scales from large scales. The best way to do so is by studying clustering in Fourier space, through the Fourier-space galaxy two-point correlation function, viz.\ the power spectrum. Hitherto, a detection of the Doppler term via power spectrum measurement has not been achieved yet, primarily because the very large scales, where relativistic effects are stronger, are afflicted by a dramatically low statistical sampling, which plagues experimental observations. To overcome this issue, \citet{2009JCAP...11..026M} proposed to use cross-correlations, due to them featuring a milder scale-dependence of the Doppler contribution.

A further complication, which we deem an engaging opportunity, stems from the fact that the relativistic effects are sample-dependent, therefore different galaxy populations display different contributions in their power spectra. For this reason, a search for tailored galaxy samples, like the one we carry out in our work, appears to be useful in the efforts to provide a detection of relativistic Doppler with the upcoming observational campaigns. As complementary case studies, we focus on a low-redshift bright-galaxy sample (BGS) and a high-redshift emission-line galaxy sample. The former is modelled after the BGS of the Dark Energy Spectroscopic Instrument \citep{2021MNRAS.507.1746E,2023arXiv230606309S}, whilst the latter mimics the H\(\alpha\) target sample of the \textit{Euclid} satellite \citep[e.g.][]{2011arXiv1110.3193L,2013LRR....16....6A,2018LRR....21....2A,2020A&A...642A.191E,2024arXiv240513491E}.

This paper is organised as follows: we introduce the auto- and cross-power spectrum with relativistic effects in \cref{sec:definitions}, outline our analysis set up in \cref{sec:metodology} and present our forecasts in \cref{sec:detection significance,sec:Fisher}---focusing on \review{a} detection significance and an information matrix approach, respectively. Then, in \cref{sec:fsky} we argue on the possibility of generalising our main results for different sky coverages and we conclude in \cref{sec:conclusions}.

\section{Definitions} \label{sec:definitions}
Up to the leading local contributions, the number density contrast of galaxy counts reads \citep{2009JCAP...11..026M,2010PhRvD..82h3508Y,2011PhRvD..84d3516C,2016JCAP...01..016D,2017PhRvD..96l3535A,2020JCAP...09..058D,2011PhRvD..84f3505B,2020JCAP...07..048B,2022JCAP...01..061C}
\begin{equation}
\label{eq:delta_g_real_space}
    \varDelta(\bm x)=b\,\delta(\bm x)-\frac{1}{\cH}\,\pd_\parallel v_\parallel(\bm x)-\alpha\,v_\parallel(\bm x)\;,
\end{equation}
with \(b\) the linear bias, \(\delta\) the matter density contrast (in comoving-synchronous gauge), \(\cH\) the conformal Hubble factor, \(\bm v\) the peculiar velocity field, and subscript `\(\parallel\)' denoting  the component of a vector along the line of sight \review{(oriented from the observer towards the source)}.
% \footnote{\review{Thus, the minus sign in front of the two line-of-sight velocity terms comes from the line of sight being opposite to the direction of incoming photons.}}
Above,
\begin{equation}
    \alpha\coloneqq\frac{\cH'}{\cH^2}+\frac2{r\,\cH}+2\,\bmag\left(1-\frac1{r\,\cH}\right)-\bevo\label{eq:AD}
\end{equation}
is the overall amplitude of the Doppler term, with a prime denoting derivation with respect to conformal time, $r$ the comoving radial distance, and \(\bmag\) and \(\bevo\), respectively, the so-called magnification and evolution bias \citep{2011PhRvD..84d3516C}.\footnote{A varied notation in the literature calls for some clarification. What we denote here by \(\alpha\) is exactly \(-A_{\rm D}\) in \citet{2021JCAP...12..009M}, \(-A\) in \citet{2020JCAP...03..065M}, and \(\alpha_{\rm GR}/(r\,\cH)\) in \citet{2017PhRvD..96l3535A}. All those definitions of the Doppler amplitude are dimensionless, contrarily to \citet{2023JCAP...04..067P} who define it as \(\alpha^{\rm(ours)}=\alpha^{\rm(theirs)}/\cH\). 
% The presence or absence of a \(1\) in the definitions just mentioned is related to either enforcing or not the validity of Euler's equation.
Regarding evolution bias, we prefer to use the letter \(\bevo\) versus other symbols used in the literature (e.g.\ \(b_{\rm e}\), \(b_{\rm evo}\), \(f_{\rm evo}\)), to reduce clutter when adding further subscripts. Finally, for magnification bias, other common notations are \(s=2\,\bmag/5\) or, globally, \(b_{\rm mag}=2-5\,s\) \citep[e.g.][see also \citealt{2021JCAP...12..009M} for a thorough review of evolution and magnification biases]{2022MNRAS.510.1964M}.} They are defined by
\begin{equation}
\label{eq:bmag_bevo}
    \bmag=-\frac{\partial \ln{n(z;L>L_{\rm c})}}{\partial \ln{L_{\rm c}}} \;, \qquad
    \bevo=-\frac{\partial \ln{n(z;L>L_{\rm c})}}{\partial \ln{(1+z)}} \;,
\end{equation}
where \(n(z;L>L_{\rm c})\) is the comoving (volumetric) number density of sources with a luminosity larger than \(L_{\rm c}\).

In Fourier space, \cref{eq:delta_g_real_space} corresponds to
\begin{equation} \label{eq:delta_g_Fourier_space}
    \varDelta(\bm k)=
    % \left(b\,+f\,\mu^2+\ii\,\frac{\cH}{k}\,\alpha\,f\,\mu\right)
    \cZ^{(1)}(\bm k)
    \,\delta(\bm k)\;,
\end{equation}
where \(\cZ^{(1)}(\bm k)=\cZ^{(1)}_{\rm N}(\bm k)+\cZ^{(1)}_{\rm GR}(\bm k)\) is the redshift-space kernel at first order in perturbation theory,
\begin{align}
    \cZ^{(1)}_{\rm N}(k,\mu)&=b\,+f\,\mu^2\;,\label{eq:Z1_N}\\
    \cZ^{(1)}_{\rm GR}(k,\mu)&=\ii\,\frac{\cH}{k}\,\alpha\,f\,\mu\;,\label{eq:Z1_GR}
\end{align}
with \(\mu\) being the cosine between the wavevector \(\bm k\) and the line of sight, and \(f\coloneqq-\de\ln\delta/\de\ln(1+z)\) the growth rate. In \cref{eq:delta_g_Fourier_space,eq:Z1_N,eq:Z1_GR}, $f\,\mu^2\,\delta(\bm k)$ is the well-known linear redshift-space distortion (RSD) term, whereas $\ii\,\cH\,\alpha\,f\,\mu\,\delta(\bm k)/k$ represents the relativistic Doppler contribution. Then, the power spectrum, i.e.\ the two-point correlation function in Fourier space, in the case of the auto-correlation of a tracer \(X\) (e.g.\ a certain galaxy population) is given by
\begin{align}
    P_{XX}(\bm k) &=\left|\cZ^{(1)}_X(\bm k)\right|^2\,P(k)\nonumber\\
    &=\left[\left(b_X\,+f\,\mu^2\right)^2\,+\,\left(\frac{\cH}{k}\,\alpha_X\,f\,\mu\right)^2\right]\,P(k)\;,\label{eq:P_auto}
\end{align}
with $P(k)$ the (linear) matter power spectrum.\footnote{The first line in \cref{eq:P_auto} comes from the fact that \(\smash{\cZ^{(1)}_{\rm N}\equiv{\rm Re}[\cZ^{(1)}]}\) and \(\smash{\cZ^{(1)}_{\rm GR}\equiv\ii\,{\rm Im}[\cZ^{(1)}]}\). Hence, \(\smash{\cZ^{(1)}(-\bm k)=[\cZ^{(1)}(\bm k)]^\ast}\), where an asterisk denotes complex conjugation.} The relativistic contribution is sub-dominant compared to the standard contributions (usually referred to as `Newtonian'), and due to its scaling $\propto k^{-2}$, relevant only on ultra-large scales.

Conversely, the situation is remarkably different for the cross-correlation between two different tracers. In general, we can write
\begin{multline} \label{eq:P_cross}
        P_{XY}(\bm k)=\cZ^{(1)}_X(\bm k)\,\cZ^{(1)}_Y(-\bm k) \, P(k)=\\
        \Bigg\{\left(b_X\,+f\,\mu^2\right)\,\left(b_Y\,+f\,\mu^2\right)+\frac{\cH^2}{k^2}\,\alpha_X\,\alpha_Y\,f^2\,\mu^2\\
        + \ii\,\frac{\cH}{k}\,\left[\alpha_X\,\left(b_Y\,+f\,\mu^2\right) - \alpha_Y\,\left(b_X\,+f\,\mu^2\right) \right]\,f\,\mu \Bigg\} \, P(k) \;,
\end{multline}
which refers to the cross-correlation power spectrum if \(X\ne Y\) and resorts to \cref{eq:P_auto} if \(X=Y\). \citet{2009JCAP...11..026M} first noted that, thanks to the not-vanishing imaginary term (which is inversely proportional to $k$), cross-correlation measurements are more promising than auto-correlations, as they allow the relativistic effects to be detected at somewhat intermediate scales. It is also interesting to stress that, in the case of cross-correlations, the real part is symmetric through the exchange of the subscripts of the tracers, i.e.\ $X\leftrightarrow Y$, whilst the imaginary part is anti-symmetric. For this reason, we can always write $P_{XY}(\bm k)=P_{YX}(-\bm k)=P_{YX}^\ast(\bm k)$.

\review{Finally, it is worth noting that our kernel in \cref{eq:delta_g_Fourier_space} only includes the leading  $\mathcal{O}(\cH/k)$ local terms and neglects further corrections as well as integrated effects \citep{2022JCAP...01..061C,2020JCAP...07..048B,2023PhRvL.131k1201F}. Other effects could be taken into account, especially lensing and the so-called wide-angle effects \citep{2018MNRAS.476.4403C,2023PhRvD.107h3528N,2023JCAP...04..067P,2024arXiv240606274J}; nevertheless, our main purpose is to coherently analyse the sample-dependent quantities and thus we leave those expansions for future works.
}

\section{Methodology} \label{sec:metodology}
The relativistic Doppler effect depends mainly on the luminosity function of the observed galaxy sample, since the evolution and the magnification biases are both defined as a logarithmic derivative of the galaxy number density (see Eq.\ \ref{eq:bmag_bevo}). Hence, we study the probability of detecting the relativistic Doppler contribution by observing two types of tracers: $\rm H \alpha$ emitters, following \citet{2021JCAP...12..009M}; and a low-$z$ bright galaxy sample (BGS), following \citep{2023AJ....165..253H,2016arXiv161100036D,2024AJ....167...62D}.

To perform a cross-correlation analysis, instead of considering all the galaxies that are observed with a flux density higher than a fixed flux cut $F_{\rm c}$ (or, equivalently, with an apparent magnitude lower than a fixed $m_{\rm c}$), we consider two complementary galaxies selections. The former is composed of all the galaxies with an observed flux $F\in[F_{\rm c},F_{\rm s})$, where $F_{\rm s}$ is the value of the flux splitting between the two samples, and the latter contains all the galaxies with $F \geq F_{\rm s}$. (Analogous for magnitudes.) Thanks to this split, we are able to obtain two independent sub-samples, faint and bright, of the same galaxy population \citep[see][]{2014PhRvD..89h3535B,2016JCAP...08..021B,2017JCAP...01..032G}. Respectively labelling the faint, bright, and total samples by subscripts `F', `B', and `T', we have $n_{\rm T}=n_{\rm F}+n_{\rm B}$, where $n_X$ is the galaxy number density for sample $X$. We choose to show in this Section how we can retrieve the galaxy, magnification, and evolution biases for the two sub-samples and leave to \ref{ap:Modelling} all the details about our way of modelling both the $\rm H \alpha$ and BGS luminosity functions.

Now, we have to calculate the different terms of \cref{eq:P_auto,eq:P_cross} for the two sub-samples. As it is shown in \citet{2014MNRAS.442.2511F}, it is possible to evaluate the linear galaxy bias, in the case of multiple samples, by means of a weighted average of the number densities of the individual biases. Hence, we can write
\begin{equation} \label{eq:Bias_faint}
    n_{\rm T}\,b_{\rm T}=n_{\rm B}\,b_{\rm B}+n_{\rm F}\,b_{\rm F}\;.
\end{equation}
As we can see, the linear bias for the faint sample is a function of $b_{\rm T}$ and $b_{\rm B}$, where in this work $b_{\rm T}$ and $b_{\rm B}$ are obtained as described in \ref{ap:Modelling}.

Concerning the magnification bias $\bmag$ and the evolution bias $\bevo$, given by \cref{eq:bmag_bevo}, we notice that the definitions of $\bmag_{\rm B}$ and $\bevo_{\rm B}$ are the same we have for the total sample, being the bright sample nothing but a total sample with a reduced sensitivity. On the other hand, for the faint sample, the expressions are somewhat different, due to the presence of the two cuts.  Specifically, in agreement with \citet{2023arXiv230604213B}, we find
\begin{align} 
    \bmag_{\rm F} &=\frac{n_{\rm T}}{n_{\rm T}-n_{\rm B}}\,\bmag_{\rm T} - \frac{n_{\rm B}}{n_{\rm T}-n_{\rm B}}\,\bmag_{\rm B} \;, \label{eq:Q_F}\\ 
    \bevo_{\rm F} &=-\frac{\partial \ln{(n_{\rm T} - n_{\rm B})}}{\partial \ln{(1+z)}}\;. \label{eq:bevo_F}
\end{align}

\begin{figure}
\centering
\includegraphics[width=\columnwidth]{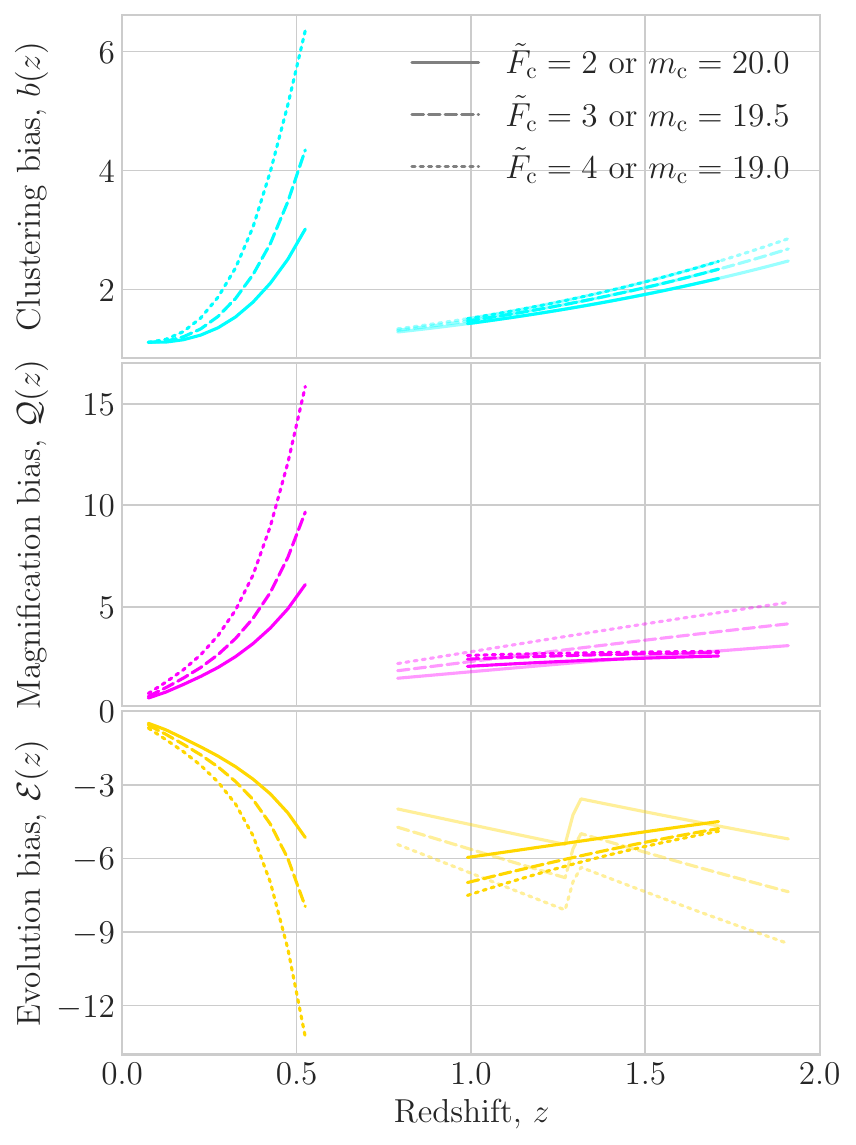}
\caption{Linear galaxy bias (cyan curves), magnification bias (magenta curves) and evolution bias (yellow curves) for all samples considered: BGS in the range \review{$z\in [0.05,0.55]$}, H\(\alpha\) \textit{Model 1} in $z\in[0.7,2.0]$, and \textit{Model 3} in $z\in [0.9,1.8]$. Different line styles correspond to different luminosity/magnitude cuts, as per the legend, where \(\tilde F=F/(10^{-16}\,\mathrm{erg\,cm^{-2}\,s^{-1}})\). (Transparency has been added to overlapping curves to the sole purpose of enhancing readability.)}
\label{fig:BGS_biases}
\end{figure}

To give the reader an idea of the properties of our samples, \cref{fig:BGS_biases} shows magnification, evolution, and clustering biases for different magnitude cuts, for both $\rm H \alpha$ galaxies and BGS. \review{Note that we model the H$\alpha$ target according to two slightly different luminosity functions, dubbed \textit{Model 1} and \textit{3} (see \ref{ap:Modelling}).} To avoid too-busy figures, we simply plot biases for different values of limiting flux/magnitudes. They correspond to the total or bright cases, depending on whether the limiting flux is the the sample's flux cut \(F_\mathrm{c}\) or our chosen splitting flux \(F_\mathrm{s}\) (or magnitude). In turn, the faint sub-sample is a function of them.

Throughout this paper, we assume a standard $\Lambda \rm CDM$ \citet{2020A&A...641A...6P} cosmology in presenting our forecasts.

\section{Detection significance} \label{sec:detection significance}
To quantify the presence of a relativistic Doppler signal in the data, and its significance, against the null hypothesis of no relativistic contributions, we rely on the \(\varDelta\chi^2\) test statistics. In our analysis, we use the theoretical predictions of \cref{eq:P_cross,eq:P_auto} to produce synthetic data, and the \(\varDelta\chi^2\) therefore corresponds simply to the chi-square for a model with no Doppler term, against our synthetic data set that includes Doppler. As it will be apparent in the next section, it is useful to introduce three dummy, binary variables: $A_{\rm N}$, $A_{\rm K}$, and $A_{\rm D}$. Respectively, they are responsible for switching off/on: the real-space clustering signal, proportional to the galaxy bias; linear RSD; and the Doppler term. Namely, we rewrite \cref{eq:Z1_N,eq:Z1_GR} as
\begin{equation}\label{eq:delta_g_with_ampl}
    \begin{split}
    \cZ^{(1)}_{\rm N}(k,\mu)&=A_{\rm N}\,b\,+A_{\rm K}\,f\,\mu^2\;,\\
    \cZ^{(1)}_{\rm GR}(k,\mu)&=\ii\,A_{\rm D}\,\frac{\cH}{k}\,\alpha\,f\,\mu\;,        
    \end{split}
\end{equation} 
Then, for a given redshift bin centred in \(\bar z_i\), the chi-square for the power spectrum reads
\begin{equation}\label{eq:significance}
    \chi_{XY}^2(\bar z_i)=\sum_{m,n} \frac{\left|P^{\rm(1,1,1)}_{XY}(k_m,\mu_n;\bar z_i)-P^{\rm(1,1,0)}_{XY}(k_m,\mu_n;\bar z_i)\right|^2}{\left[\varDelta P_{XY}^{\rm(1,1,1)}(k_m,\mu_n;\bar z_i)\right]^2}\;,
\end{equation}
where we assume, as customary, that the covariance matrix is diagonal in \(k\)-, \(\mu\)-, and \(z\)-space, having denoted by \(\varDelta P_{XY}^2\) the variance on a measurement of \(P_{XY}\). In the expression above, superscripts in parentheses refer to the values of $(A_{\rm N},A_{\rm K},A_{\rm D})$. The variance is computed for the total signal, including Doppler. (Note that this does not affect the results significantly, as Doppler is a sub-dominant term.)

Assuming a Gaussian covariance matrix for the power spectrum signal---accurate enough on the large, linear scales we are interested in---the variance associated with a measurement of $P_{XY}(\bm k)$ averaged in a given redshift bin reads
\begin{equation}
    \varDelta P_{XY}^2(\bm k)=\frac{1}{N_{\bm k}}\,\left[\tilde P_{XY}(\bm k)\,\tilde P_{YX}(\bm k)+\tilde P_{XX}(\bm k)\,\tilde P_{YY}(\bm k)\right]\;,\label{eq:variance}
\end{equation}
having defined $\tilde P=P+N$, with $N$ the noise related to a measurement of $P$, and $N_{\bm k}=2\,\pi\,k^2\,\varDelta k\,\varDelta\mu/k_{\rm f}^3$. The latter quantity represents the number of independent modes available in the observed volume \(V\), having explicited the fundamental frequency \(k_{\rm f}=2\,\pi\,V^{-1/3}\). Lastly, \(\varDelta k\) and \(\varDelta\mu\) denote the sizes of the \((k,\mu)\)-bins, whereas \(V\) depends on the redshift width, \(\varDelta z\).

In the case of the galaxy power spectrum, the noise power spectrum is a scale-independent, shot-noise term due to galaxies discretely sampling the underlying continuous matter distribution. Specifically, we have \(\smash{N_{XY}(\bm k;\bar z_i)\equiv N_{XY}(\bar z_i)=\delta_{\rm(K)}^{XY}/\bar n_X(\bar z_i)}\), with \(\delta_{\rm(K)}\) the Kronecker delta symbol and \(\bar n_X\) the mean (volumetric) galaxy number density. Finally, it is worth noting that the variance of the faint-bright cross-correlation power spectrum, $P_{\rm FB}$, is real, despite the signal itself being complex. This is due to the fact that \review{\(P_{XY}=P_{YX}^\ast\)} for any pair of tracers.

In our analysis, we fix the largest wavenumber, \(k_{\rm max}\), to \(\smash{k_{\rm nl}=0.2\,h\,\mathrm{Mpc}^{-1}}\), viz.\ the scale at which non-linear effects take over the linear growth of structure at \(z=0\). This is a conservative approach, since \(k_{\rm nl}\) is a redshift-dependent quantity, monotonically increasing with redshift; and, furthermore, there are available recipes to push at least to mildly non-linear scales. But our choice is motivated by the Doppler signal scaling with \(k^{-1}\) or \(k^{-2}\), meaning that large wavenumbers will in practice not contribute to the detection of the relativistic signal. On the other hand, the smallest wavenumber, $k_{\rm min}$, is a critical quantity. We choose to fix it to the fundamental frequency, \(k_{\rm f}\). By definition, this quantity is survey-dependent, being determined by the redshift range and binning, and by the observed fraction of the sky, \(f_{\rm sky}\). For the sake of generality, we shall for now assume to be observing the entire sky, providing later on the reader with means to rescale our findings to any \(f_{\rm sky}\).

For the definition of the redshift ranges we report the details in \ref{ap:Modelling}. Since we are interested in the largest scales, we choose to take slightly thicker redshift bins than what used in most galaxy survey forecasts, with $\varDelta z \approx 0.2$---but we shall also discuss different binning choices. For $\rm H \alpha$ emitters, our choice implies a total of \(7\) \(z\)-bins for \textit{Model 1} and \(5\) \(z\)-bins for \textit{Model 3}. On the other hand, for the BGS, we have \(3\) redshift bins. Finally, we adopt \(30\) log-spaced \(k\)-bins in the range \(k\in[k_{\rm min},k_{\rm max}]\) and \(10\) \(\mu\)-bins in the range \(\mu\in[-1,1]\). We have checked that the actual number of \(k\)- and \(\mu\)-bins does not significantly affect the final results.
\begin{figure*}
\centering
\includegraphics[width=\textwidth]{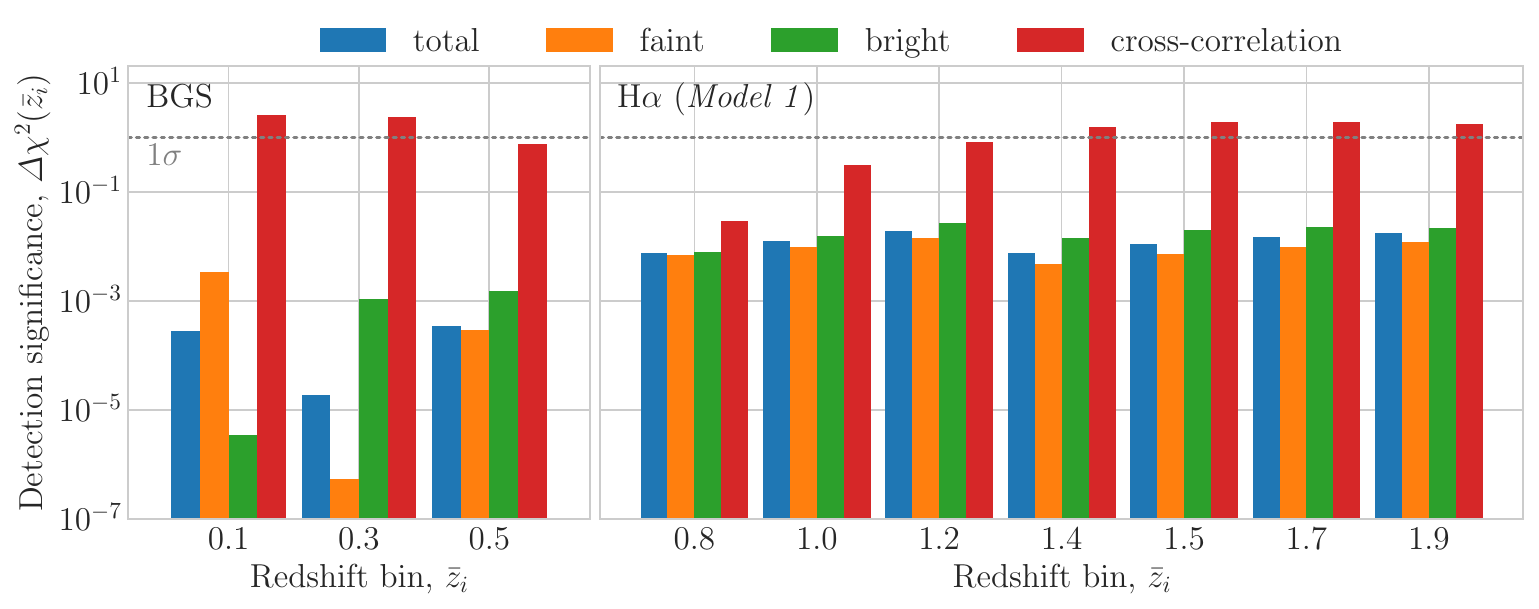}
\caption{Detection significance of relativistic Doppler contribution for an $\rm H \alpha$ \textit{Model 1} luminosity function (right, high-$z$ panel) and the BGS (left, low-$z$ panel). The $\varDelta\chi^2$ variable is evaluated against a null hypothesis of no Doppler contribution in the galaxy power spectrum. Colour code: blue, orange, and green respectively for auto-correlation power spectrum of total, faint, and bright samples; red is for the faint-bright cross-correlation power spectrum.}
\label{fig:detection-significance-z}
\end{figure*}

Illustratively, in the right-hand panel of \cref{fig:detection-significance-z} we display the detection significance for the relativistic Doppler term for the \textit{Model 1} $\rm H \alpha$ luminosity function. We consider two galaxy populations with \(\smash{F_{\rm c}=2.0\times10^{-16}\,\mathrm{erg\,cm^{-2}\,s^{-1}}}\) and splitting flux between the two samples \(\smash{F_{\rm s}=3.2\times10^{-16}\,\mathrm{erg\,cm^{-2}\,s^{-1}}}\). The value of $F_{\rm s}$ is chosen so that the faint and bright samples are almost equally dense. Analogously, we show in the left-hand panel the differential statistics significance for the BGS with $m_{\rm c}=20.175$ and $m_{\rm s}=19.5$.

First of all, by comparing separately the orange (faint sample) and the green (bright sample) histograms to the blue (total sample) one, it is clear that a larger statistics---that is, the total sample, with the largest number density---does not imply a better measurement of the Doppler term in either tracer, be it $\rm H \alpha$ emission-line galaxies or bright galaxies. This happens because on the largest scales, where the Doppler contribution becomes non-negligible, shot-noise is not an issue. On the other hand, the larger \(\bevo\) and \(\bmag\) (in absolute value), the larger the amplitude of the Doppler term \(\alpha\) (see Eq.~\ref{eq:AD}), and for both sub-samples the improvement of \review{\(\alpha\)} roughly compensates for the sparsity of the samples. However, the average \(\varDelta\chi^2\) per redshift bin remains at most of order \(0.01/0.001\) for \(\rm H \alpha\)/BGS, making any detection effectively unrealistic. Conversely, the use of the cross-correlation turns the situation around, thanks to the appearance of the \(\propto k^{-1}\) term in \cref{eq:P_cross}. Indeed, per-bin \(\varDelta\chi^2\) values are now well above unity for almost all the redshifts considered, with an average improvement of a factor \(100\)--\(1000\) over the auto-correlation power spectrum of the total sample. This demonstrates the constraining power of the cross-power spectrum and the luminosity cut technique. This is in agreement with \citet{2023arXiv230604213B}, who looked at the dipole of the galaxy two-point correlation function in real space as a proxy of the Doppler term \citep[see also][for the cross-correlation dipole, mainly at small-scales]{2020MNRAS.498..981S,2022MNRAS.511.2732S,2023MNRAS.524.4472S}.

\begin{figure}
\centering
\includegraphics[width=\columnwidth]{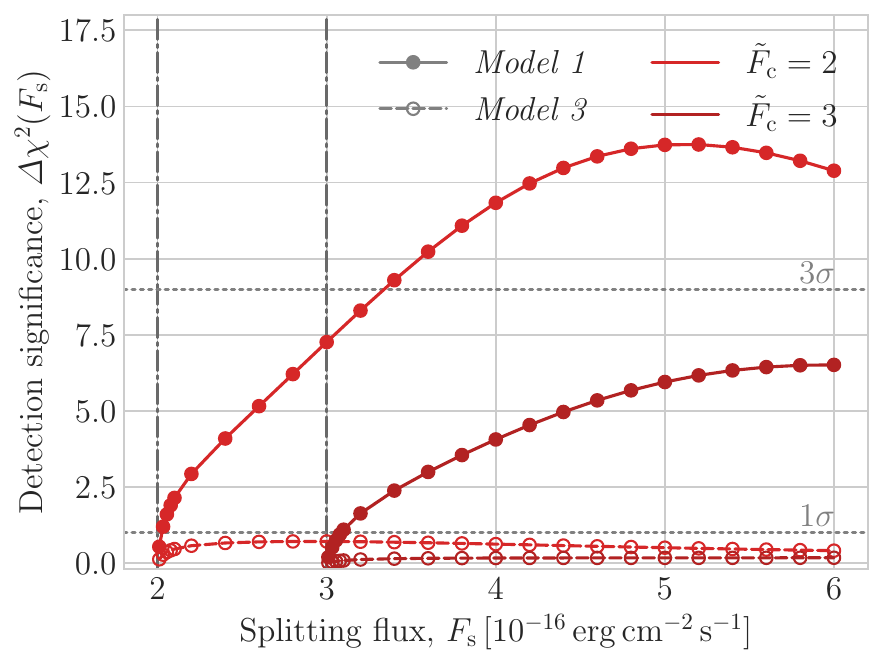}
\caption{Cumulative statistical significance of the detection of the relativistic Doppler effect associated with a faint-bright cross-correlation power spectrum measurement as a function of $F_{\rm s}$\review{, for a H$\alpha$ galaxy population}. Solid curve is for \textit{Model 1} and dashed curve for \textit{Model 3}, whilst dotted lines mark $1\sigma$, $3\sigma$, and $5\sigma$ significance levels. Results for two flux cuts are depicted, i.e.\ $\tilde F_{\rm c}=2.0$ (red) and $\tilde F_{\rm c}=3.0$ (dark red), with \(\tilde F=F/(10^{-16}\,\mathrm{erg\,cm^{-2}\,s^{-1}})\).}
\label{fig:Euclid_detection-significance}
\end{figure}

\begin{figure}
\centering
\includegraphics[width=\columnwidth]{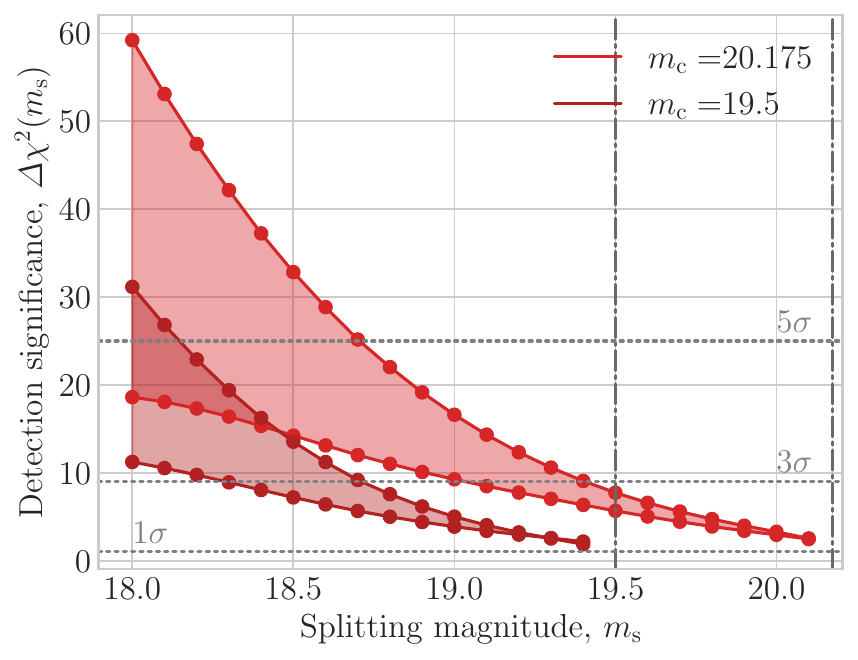}
\caption{Same as \cref{fig:Euclid_detection-significance}, but in the case of BGS. Shaded areas around each curve bracket the dependence upon \(\varDelta z\), from \(\sim0.17\) (bottom edge) to \(\sim0.5\) (top edge). Results are plotted as a function of $r$-magnitude, for two magnitude limits, namely, $m_{\rm c}=20.175$ (light-red) and $m_{\rm c}=19.5$ (dark red).}
\label{fig:BGS_detection-significance}
\end{figure}

Now, focusing on the most promising observable---the cross-correlation power spectrum---\cref{fig:Euclid_detection-significance,fig:BGS_detection-significance} show the total $\varDelta\chi^2$ associated with the detection of the relativistic Doppler effect, cumulative over all redshift bins, as a function of the splitting flux/magnitude, $F_{\rm s}$ or $m_{\rm s}$, for two values of survey sensitivity, $F_{\rm c}$ or $m_{\rm c}$. In particular, the former refers to a \textit{Euclid}/\textit{Roman}-like H\(\alpha\) emission-line galaxy survey, whereas the latter to a DESI-like bright galaxy survey.

Let us start from \cref{fig:Euclid_detection-significance}, where both \textit{Model 1} (solid curves) and \textit{Model 3} (dashed curves) are presented. The reference \(\varDelta z\approx0.2\) is shown, but we have also checked that varying the redshift-bin width within the range \(\varDelta z\approx0.05\) to \(0.3\) does not lead to significant differences in the results. In the reference scenario, for the higher detector sensitivity case, i.e.\ with the lower $F_{\rm c}$ (denoted with a slightly brighter red colour), the cumulative $\smash{\sqrt{\varDelta\chi^2}}$ for \textit{Model 1} reaches readings above the \(3\,\sigma\) detection threshold when $F_{\rm s}\geq3.5\times 10^{-16}\, \rm erg\,cm^{-2}\,s^{-1}$. On the other hand, results obtained with \(F_{\rm c}=3.0\times10^{-16}\,\mathrm{erg\,cm^{-2}\,s^{-1}}\) (darker red curves and markers) show a similar behaviour, reaching for \textit{Model 1} \(1\,\sigma\) confidence level if $F_{\rm s}\geq3.1\times 10^{-16}\, \rm erg\,cm^{-2}\,s^{-1}$. Conversely, \textit{Model 3} curves do not look that promising, as they do not allow for a detection of the relativistic contribution, even though they are much higher than those for the corresponding auto-correlation measurements (not shown).

\review{Despite the fact that difference between the two models of the luminosity function is partly driven by the number densities, by comparing results from \textit{Model 1} at $F_{\rm c}=3.0\fluxum$ and \textit{Model 3} at $F_{\rm c}=2.0\fluxum$, it is remarkably clear that $\alpha$ is the major responsible for this outcome. Indeed, we have checked that the impact of the variation of the available $z$-range is minimal, with the curve reduced only by a factor of about $10\%$ if we run the analysis on \textit{Model 1} using the redshift range of \textit{Model 3}. Also, by comparing the results of \textit{Model 1} at the worse $F_{\rm c}$ with those of \textit{Model 3} with the optimistic flux cut, we see that the difference between the two outcomes is still clearly noticeable, even though the number densities (and thus the shot noises) are similar.} These findings remind us of the importance of modelling the relativistic sample-dependent biases properly. Since, when we deal with relativistic corrections, different galaxy populations display different contributions in their power spectra, we have to face the issue of describing their luminosity function features correctly if we aim to study the probability of detecting the relativistic Doppler, as was clearly stated in \citet{2020JCAP...03..065M,2021JCAP...12..009M}.

\Cref{fig:BGS_detection-significance} depicts the cumulative detection significance for the DESI-like BGS, as a function of the splitting $r$-magnitude, for two threshold values $m_{\rm c}=20.175$ and $m_{\rm c}=19.5$---corresponding respectively to the so-called BGS bright and BGS faint \citep{2023AJ....165..253H}. In this case, we choose to draw the \(\varDelta z\approx0.5\) (upper edges of the shaded areas) results---that is, the 1 $z$-bin result---on top of the reference \(\varDelta z\approx0.2\) curves (lower lines). As expected, the larger the redshift bins, the higher the $\varDelta\chi^2$, due to the increase in the volume, which is particularly relevant at very low redshift. Interestingly, we reach a $3\,\sigma$ detection in the reference scenario for $m_{\rm s}\leq19.1(18.3)$ with $m_{\rm c}\leq20.175(19.5)$ whereas we are able to obtain a detection well above $5\,\sigma$ confidence level if we consider only one redshift bin---namely, we find readings $\varDelta\chi^2\geq 25$ where $m_{\rm s}\leq18.7(18.2)$ and $m_{\rm c}\leq20.175(19.5)$. We conclude, by comparing \cref{fig:Euclid_detection-significance,fig:BGS_detection-significance}, that BGS turns out to be more likely to allow us to observe the relativistic Doppler effect in the future than $\rm H \alpha$ sample.

\review{This finding is somehow expected, since there are claims in the literature of the Doppler being dominant at low redshift and overtaken by the other corrections as $z$ increases. However, we argue that in our analysis this outcome is mainly driven by sample specifications, i.e.\ the differences in the Doppler amplitude due to $\bmag$ and $\bevo$, rather than follow from a way general perspective \citep[see e.g.\ ][for an example of high-$z$ Lyman-break galaxy populations]{2024arXiv240305398M}. Furthermore, the signal we refer to as relativistic Doppler is mainly given by the imaginary term within the cross-correlation, which in fact results in a dipole, hence it is not directly comparable with monopole power spectra shown, for instance, in \citet{2022JCAP...01..061C}.}

\section{Parameter constraints} \label{sec:Fisher}
We now move to estimate the error associated with a measurement of each of the contributions to the power spectrum, namely the three parameters we introduced before: the amplitude of the Newtonian term, $A_{\rm N}$; that of the Kaiser RSD contribution, $A_{\rm K}$; and that of the Doppler effect, $A_{\rm D}$. Going back to \cref{eq:delta_g_Fourier_space}, we include these dummy variables as amplitude parameters (see Eq.\ \ref{eq:delta_g_with_ampl}), whose fiducial values are fixed to $A_{\rm N}=A_{\rm K}=A_{\rm D}=1$. Then, the uncertainty on a measurement of them can be evaluated through an information matrix analysis. In the \(i\)th redshift bin, the information matrix for the parameter set ${\bm\theta}=\{A_{\rm N},A_{\rm K},A_{\rm D}\}$ reads
\begin{multline}\label{eq:fisher}
    {I}_{\alpha\beta}(\bar z_i)=\displaystyle \sum_{m,n} \left[\varDelta P_{XY}^{\rm(1,1,1)}(k_m,\mu_n;\bar z_i)\right]^{-2}\\
    \times\left[\frac{\partial P_{XY}(k_m,\mu_n;\bar z_i)}{\partial \theta_{(\alpha}}\,\frac{\partial P_{XY}^\ast(k_m,\mu_n;\bar z_i)}{\partial \theta_{\beta)}}\right]_{\bm\theta=1}\;,
\end{multline}
where parentheses around indexes denote symmetrisation, the variance is again given by \cref{eq:variance}, and the sum runs over all the configurations $(k_m,\mu_n)$. Hence, the total Fisher matrix, cumulative over all redshift bins, is simply \({\sf I}=\sum_i{\sf I}(\bar z_i)\). Lastly, the cumulative marginal errors on \(\{\theta_\alpha\}\) are given by $\sigma_{\theta_\alpha}=\smash{\sqrt{({\sf I}^{-1})_{\alpha \alpha}}}$. It is useful to stress, at this point, that the definition of the information matrix \cref{eq:fisher} is consistent for both auto- and cross-correlations. Concerning auto-correlations ($X=Y$), the symmetrisation is in fact trivial, just because the power spectrum is real. On the other hand, symmetrisation plays a crucial role in the case of cross-correlation $X \neq Y$. Being $P_{XY}$ complex, it ensures the information matrix to be real.

Since the differential detection significance analysis points out the supremacy of the faint-bright cross-correlation, in this section we focus on this case alone, leaving aside auto-correlation forecasts.
\begin{figure}
\centering
\includegraphics[width=\columnwidth]{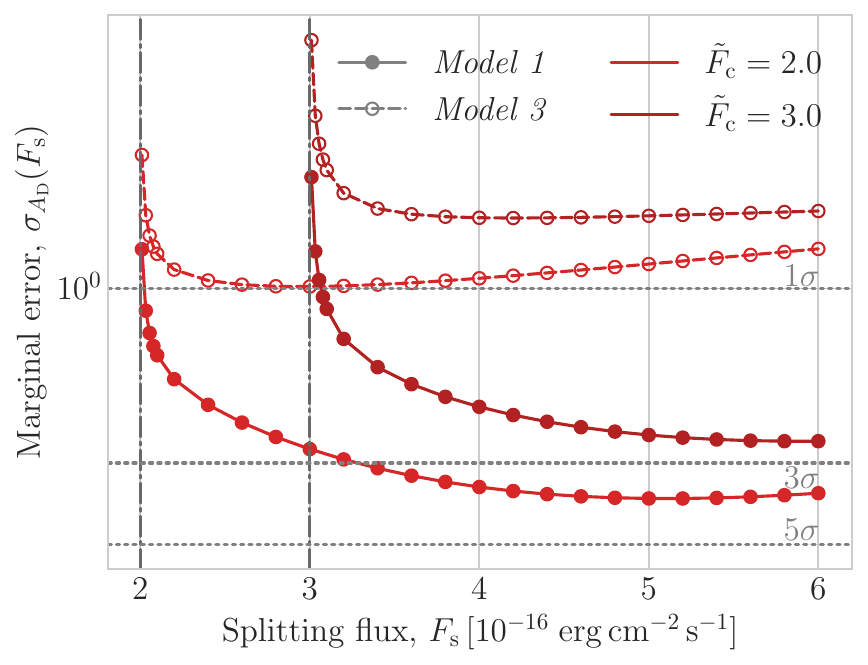}
\includegraphics[width=\columnwidth]{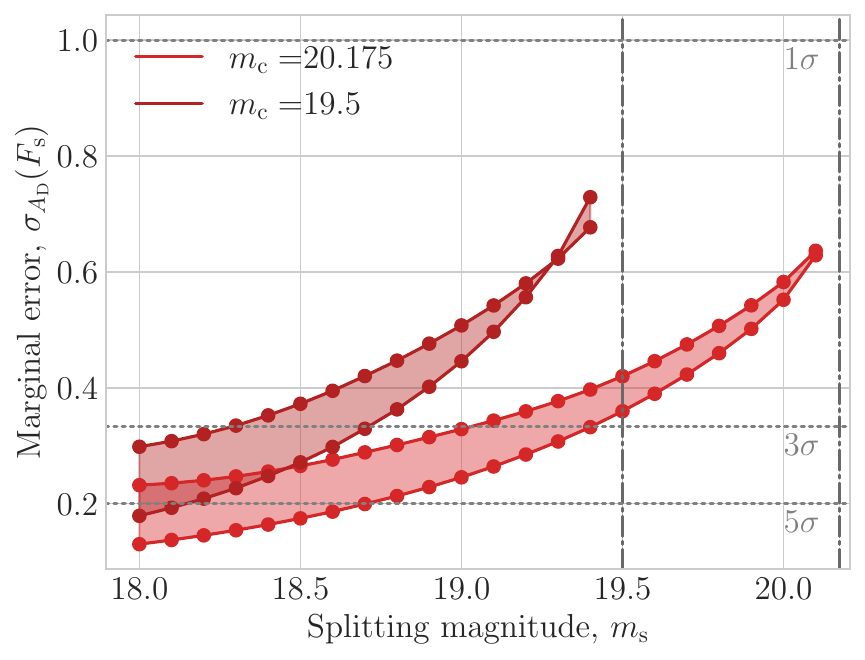}
\caption{Cumulative marginal error on a measurement of the Doppler amplitude $\sigma_{A_{\rm D}}$ as a function of the adopted split (as in \cref{fig:Euclid_detection-significance,fig:BGS_detection-significance} we use a flux-based description for $\rm H \alpha$ emiterrs and a $r$-band magnitude-based framework for BGS). Top panel: $\rm H \alpha$ galaxies, \textit{Model 1} and \textit{3} are shown with solid and dashed lines, respectively. Bottom panel: DESI-like BGS results, curves for $\varDelta z \sim 0.17$ and $\varDelta z \sim 0.5$ are drawn. Note that limits of shaded areas are now reversed, with \(\varDelta z\simeq0.17(0.5)\) being the top(bottom) edge.}
\label{fig:sigmaAD}
\end{figure}
Analogously to \cref{fig:Euclid_detection-significance,fig:BGS_detection-significance}, \cref{fig:sigmaAD} shows, for both galaxy populations, the cumulative marginal error on the estimation of \(A_{\rm D}\), $\sigma_{A_{\rm D}}$, as a function of the adopted flux/magnitude split. In the top panel, which refers to $\rm H \alpha$ emission-line galaxies, solid and dashed lines are for \textit{Model 1} and \textit{Model 3}, respectively. The lower panel shows $\sigma_{A_{\rm D}}(m_{\rm s})$ for the BGS, for either a wide or a narrow $z$-bins scenario, as before. Since the fiducial value is set to $A_{\rm D}=1$, we at least need $\sigma_{A_{\rm D}}<1$ in order to claim a detection of the relativistic contribution. For \textit{Model 1} this condition is almost always verified, whereas for \textit{Model 3} it is never. On the other hand, BGS always has a marginal error lower than $\sigma_{A_{\rm D}}=1$.

It is worth noting that these findings on the behaviour of the galaxy samples further confirm the cumulative detection significance results. \review{It may be noticed that the constrain coming from the thick-bin scenario becomes slightly worse than that in the case $\varDelta z \simeq 0.17$ when $m_{\rm c}=19.5$ and $m_{\rm s}=19.4$. Despite this is a minimal difference and does not modify our general discussion, we point out that such behaviour might arise out of our assumption of evaluating all the quantities at $\bar z_i$, that is, the central value of the $z$-bins, so that we might have lost some accuracy in the study of the single bin case.}

Furthermore, we find a clear relation between the two statistical tools we adopt: because of the $A_{\rm D}$ definition, we have $\smash{\sqrt{\varDelta \chi^2 }\sim \sigma_{A_{\rm D}}^{-1}}$. Indeed, $A_{\rm D}$ is almost uncorrelated to either $A_{\rm N}$ and $A_{\rm K}$, due to its presence in the imaginary part of the spectrum \citep{2009JCAP...11..026M}. To estimate the correlation between two parameters we can use the elements of the information matrix, defining the correlation coefficient (no implicit summation)
\begin{equation}
    \rho_{\alpha\beta}=\frac{\big({\sf I}^{-1}\big)_{\alpha\beta}}{\sqrt{\big({\sf I}^{-1}\big)_{\alpha\alpha}\,\big({\sf I}^{-1}\big)_{\beta\beta}}}\;,
\end{equation}
with $\theta_\alpha$ and $\theta_\beta$ the two parameters under consideration. If we compute the correlation coefficients between $A_{\rm D}$ and either $A_{\rm N}$ or $A_{\rm K}$ in the case of the faint-bright correlation we always find values $<21\%$, which clearly mean that \(A_{\rm D}\) is basically uncorrelated with either two of the other parameters, which are instead quite correlated between themselves.

\begin{figure}
    \centering
    \includegraphics[width=\columnwidth]{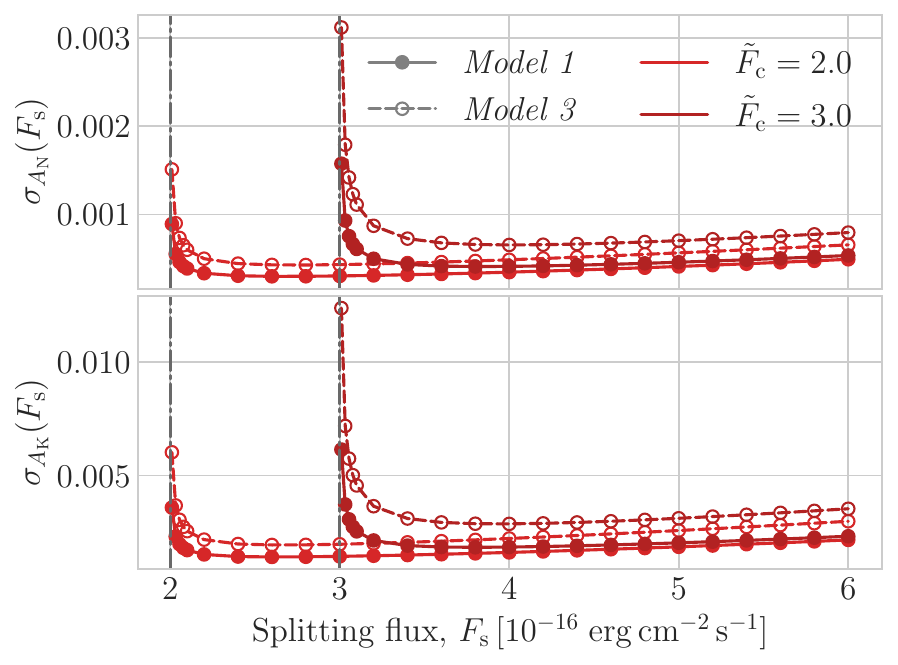}
    \includegraphics[width=\columnwidth]{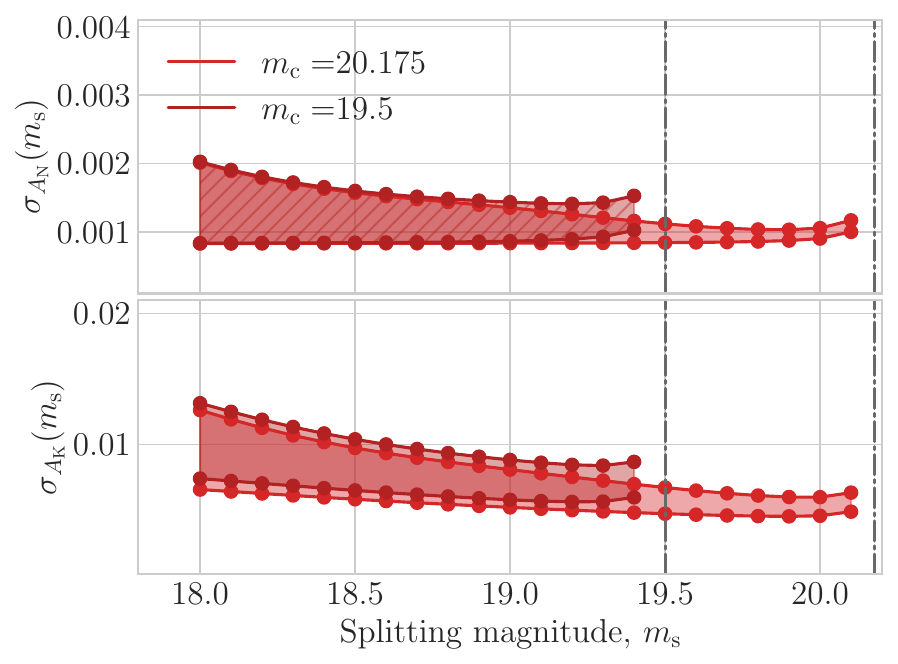}
    \caption{Cumulative marginal error on the estimation of $A_{\rm N}$ and $A_{\rm K}$ as a function of the splitting flux or $r$-magnitude, for both $\rm H \alpha$ emitters (top panel) and DESI-like BGS (bottom panel), for the faint-bright cross-correlation. In each panel, the upper part concerns the Newtonian amplitude, while the lower one is for the Kaiser RSD term. Above, solid and dashed lines are for \textit{Model 1} and \textit{Model 3}, respectively; below, shaded areas bracket $0.17 \lesssim \varDelta z \lesssim 0.5$. Light-red curves refer to the optimistic cut ($F_{\rm c}=2.0\fluxum$ or $m_{\rm c}=20.175$), whilst dark-red ones are for $F_{\rm c}=3.0\fluxum$ or $m_{\rm c}=19.5$.}
    \label{fig:sigmaANAK}
\end{figure}
\Cref{fig:sigmaANAK} shows the results of the information matrix analysis for $\sigma_{A_{\rm N}}$ and $\sigma_{A_{\rm K}}$. As before, forecasts are presented for both an optimistic luminosity threshold---i.e.\ $F_{\rm c}=2.0\fluxum$ or $m_{\rm c}=20.175$, light-red curves---and a pessimistic one---that is, $F_{\rm c}=3.0\fluxum$ or $m_{\rm c}=19.5$, dark-red lines. 
First of all, we note that, being the Newtonian and the Kaiser contributions dominant, the constraining power on $A_{\rm N}$ and $A_{\rm K}$ of the cross-correlation power spectrum is much stronger than that for the relativistic contribution. In the case of $\rm H \alpha$ galaxies, almost all of the probed splitting fluxes---namely $F_{\rm s} \in [2.0,6.0]\fluxum$---correspond to tiny marginalised errors, such that $\sigma_{A_{\rm N}}<1.0 \times 10^{-3}$ and $\sigma_{A_{\rm K}}<4.0 \times 10^{-3}$. Analogously, we have $\sigma_{A_{\rm N}}<2.0 \times 10^{-3}$ and $\sigma_{A_{\rm K}}<1.5 \times 10^{-2}$ when $m_{\rm s} \in [18.0,20.1]$ for DESI BGS.
In addition, as expected on the basis of the improvement in the galaxy survey sensitivity, marginalised errors associated with a measurement of the dominant terms with the optimistic flux cuts are better than those with the more conservative limits. Such a behaviour looks reasonable because of the higher statistics achieved by a survey with a lower(higher) flux(magnitude) limit. 
Even the shape of the curves $\sigma_{A_{\rm N}}$ and $\sigma_{A_{\rm K}}$ as a function of the flux/magnitude split can be easily explained: the presence of a minimum in all the studied configurations means that there exists an optimal value of the splitting flux that can maximise the detection probability of the dominant terms. 

Also, we discuss the impact of diverse redshift bins. We notice that whichever width $\varDelta z \in [0.05,0.3]$ almost leads to the same information matrix results for $\rm H \alpha$ emitters. Conversely, because of the greater relative gain in the volume at low-$z$, we find an appreciable improvement in the marginal errors in moving from a narrow ($\varDelta z \sim 0.17$) to a wide ($\varDelta z = 0.5$) $z$-bins scenario, as pointed out in \cref{sec:detection significance}. However, in each considered case the variations due to statistical choices appear to be far smaller than the differences due to the physical observable---that is, auto- versus cross-correlation---or to the choice of the galaxy samples.
This test conveys an important message: it suggests that the specific statistical framework adopted in our analysis does not significantly affect the results. In such a context, the luminosity cut technique looks promising also because it proves to be solid, in the sense that different data-analysis approaches lead to essentially the same results.

\section{Partial sky coverage}\label{sec:fsky}
Let us remind the reader that the results hitherto presented refer to the idealistic case of full-sky observations. In truth, this is never the case, either because a ground-based telescope will have access only to a portion of the celestial sphere, or because even a space-born experiment will not be able to pierce through our own Galaxy, whose stars will block the line of sight and whose diffuse gas will absorb incoming radiation. Nonetheless, for the wide sky coverages usually envisioned for cosmological analyses, the effect of partial sky can be folded in by rescaling the volume \(V\) in \cref{eq:variance} by the fraction of observed sky, \(f_{\rm sky}\).

For most applications, the effect of an $f_{\rm sky}\ne1$ on a measurement of \(P_{XY}\) can be thought of as a simple inflation of the error bars by a factor \(\smash{f_{\rm sky}^{-1/2}}\). However, in the present case in which we seek to detect an effect relevant on the largest scales, we should also in principle account for the fact that a smaller volume directly translates into a larger fundamental frequency, \(k_{\rm f}\); and that the fundamental frequency is the smallest wavenumber observable. Hence, any $f_{\rm sky}\ne1$ would call for a rerun of the analysis, with a larger \(k_{\rm min}\) (which, we remind the reader, we fix equal to \(k_{\rm f}\)) and, possibly, a different \(k\)-binning. However, we find that this effect is relevant only when looking at the differential, per-bin \(\varDelta\chi^2\) and \(\sigma_{A_{\rm D}}\), whilst it effectively cancels out in the cumulative ones. More quantitatively, the simple rescaling of the variance of \cref{eq:variance} by \(f_{\rm sky}\) works remarkably well, with discrepancies \(<15\%\) even for sky coverages as small as \(f_{\rm sky}\simeq0.05\) (except for the anyways non-detectable case of \textit{Model 3} with $F_{\rm c}=3.0\fluxum$, where we find discrepancies up to $\sim 40\%$). Thanks to the validity of such scaling relation, our main results of \cref{fig:Euclid_detection-significance,fig:BGS_detection-significance,fig:sigmaAD} can be easily related to any \(f_{\rm sky}\).

\section{Discussion and conclusions}\label{sec:conclusions}
A measurement of a peculiar GR effect on cosmological scales would be an astonishing confirmation of the validity of Einstein's theory in a regime where it is poorly probed experimentally. With the upcoming galaxy surveys, this wish looks set to become a reality, thanks to the unprecedented cosmic volumes probed, which will allow us to sample even the largest scales in the structure of the Universe \citep[see e.g.][]{2020JCAP...07..048B}. In this work, we have presented forecasts for the detection of the relativistic Doppler term with galaxy power spectrum measurements, for both auto- and cross-correlations of various galaxy samples. Since the amplitude of such Doppler term differs according to the target galaxy population, we set to the task of optimising sample selection, for the search for a relativistic signature. In particular, we have focused on two complementary tracers of the cosmic large-scale structure: a sample of bright galaxies at low redshift and an higher-redshift sample of emission-line galaxies. This is done in the spirit of the oncoming data from DESI and the \textit{Euclid} satellite.

The contribution from the dominant relativistic Doppler term, is to be relevant only on very large scales in the case of auto-correlation measurements, but its presence in the imaginary part of the cross-power spectrum appears to be measurable even at somewhat intermediate scales \citep{2009JCAP...11..026M}. Thus, the comparison between auto- and cross-correlation measurements points out the supremacy of the latter, mostly due to the milder scale dependence of the Doppler term. When the cross-correlation is performed over two non-overlapping sub-samples of faint and bright galaxies, the differential Doppler detection significance is around two orders of magnitude larger over the entire redshift range than that found in the auto-correlation cases.

Considering the whole redshift range, i.e.\ the cumulative cross-correlation detection significance, with a DESI-like BGS we can obtain a detection of the relativistic Doppler effect with a confidence level well above $3\,\sigma$, by carefully selecting $m_{\rm s}$, namely the value of the $r$-magnitude that splits the two sub-samples. However, this is not the case for $\rm H \alpha$ emitters, where we have tested two luminosity function models but only one of them seems to allow for a detection. As a consequence, we state that in the present case, a bright galaxy survey (like DESI BGS) is a better target to look at. This finding does not come as unexpected, since relativistic Doppler should be more relevant at low redshift. Quantitatively, the $3\,\sigma$ level is reached in a DESI-like BGS in the case of a maximum $r$-band magnitude of $m_{\rm c}=20.175(19.5)$ whether \(m_{\rm s}<19.4\)--\(19.0(18.7\)--\(18.3)\), depending on the redshift bin width. On the $\rm H \alpha$ side, only the \textit{Model 1} curve with $F_{\rm c}=2.0\fluxum$ reaches the $3 \,\sigma$ detection, when \review{$F_{\rm s}>3.4\fluxum$}.

Results for the detection significance are confirmed by the estimation of the marginal error via the information matrix formalism. The marginal error associated with a measurement of the Doppler contribution has a minimum at about $F_{\rm s}\sim F_{\rm c} + 2.5\fluxum$ for $\rm H \alpha$ \textit{Model 1} and $F_{\rm s}\sim F_{\rm c} + 0.5\fluxum$ for \textit{Model 3}. The presence of such a minimum is meaningful, since it tells us we can somehow fine-tune the faint-bright sub-division to maximise the probability of detecting the GR effect. Conversely, the Doppler cumulative marginal error for a DESI-like BGS improves as we depopulate the bright sample, although we would have expected to find a minimum. From a less theoretical point of view, we cannot assume the marginal error to be decreasing monotonically as we push up (in luminosity) the split between the two sub-samples, because, at a certain point, the bright sample should become too sparsely populated to extract information. It is therefore worth noting that the choice of the optimal splitting flux (or magnitude) has to take into account at least also the behaviour of the marginal errors on the dominant contributions---namely, the clustering (Newtonian) and RSD (Kaiser) terms. Indeed, these in principle might show a different position of the minimum, as is the case, hence we cannot forget about them, even in a purely theoretical study.

In addition, a check on the impact of the statistical framework used---that is the width of the bins in $z$, $\mu$, $k$---tells us that all the analyses presented are robust, in the sense that the results do not vary significantly if we vary the analysis set-up.

To conclude, the luminosity cut technique proposed by \citet{2014PhRvD..89h3535B} appears to be very promising for power spectrum analyses, not only because it makes cross-correlation measurements possible using just one data set, but also because for specific faint-bright divisions, it can somehow boost the relativistic contribution. Note that the results presented here come, for the first time, from a fully self-consistent treatment of all the relevant observational quantities, such as the galaxy bias and the magnification and evolution biases. We have been able to achieve this thanks to the implementation of analytical scaling relations calibrated on real data \citep[see also][]{2020MNRAS.495.1340F}.

\review{Interesting extensions of this work are going to include the full relativistic correction to the galaxy number density in a power spectrum analysis in harmonic space \citep{TesiMarco}, with the additional purpose of better investigating whether the Doppler contribution is effectively the dominant correction. Furthermore, a theoretical study will have to assess the non-trivial interplay between wide-angle and relativistic effects when using the luminosity cut technique.}
Future studies will \review{also} have to assess the reliability of the strategy using simulated data. For instance, it will be important to assess how much some aspects are due to the modelling adopted. Moreover, flux density measurements are subject to several technical details, like the angular size and the magnitude of the target galaxy, the exposure time, etc. For this reason, some issues might occur in the definition of the sub-samples, and thus caution might be needed in the choice of the splitting flux and in the estimation of the galaxy brightness uncertainty. 

\section*{Acknowledgments}
The authors acknowledge support from the Italian Ministry of University and Research (\textsc{mur}) through PRIN 2022 `EXSKALIBUR – Euclid-Cross-SKA: Likelihood Inference Building for Universe's Research', Grant No.\ 20222BBYB9, CUP C53D2300131 0006, from the Italian Ministry of Foreign Affairs and International Cooperation (\textsc{maeci}), Grant No.\ ZA23GR03, and from the European Union -- Next Generation EU.  They also thank C.\ Clarkson and R.\ Maartens for their reading of the final draft of the manuscript, as well as C.\ Hahn and C.\ Yeche for advice on BGS bias estimations, and N.\ Fornengo, for useful questions raised during the presentation of earlier stages of FM's thesis, whence this work stemmed.

\appendix \section{Modelling galaxy populations} \label{ap:Modelling}
\subsection{$\rm H \alpha$ survey}
To derive all the key quantities in a self-consistent way for $\rm H \alpha$ galaxy populations, we start from the galaxy luminosity function. In this work, we use models that present a factorised form for the luminosity function, 
\begin{equation}\label{eq:lumfunc_factorised}
    \phi(z,L)=\phi_\ast(z)\,g[y(z,L)]\;,
\end{equation}
where $\phi_\ast$ is a characteristic number density and $y(z,L)=L/L_\ast(z)$ is the luminosity $L$ normalised with respect to a characteristic luminosity $L_\ast(z)$. We adopt two models for the luminosity function: one of Schechter type and another coming from fits to observational data. In line with the nomenclature of \citet{2016A&A...590A...3P}, we dub them \textit{Model 1} and \textit{Model 3}, respectively. For the definition of the redshift ranges for \textit{Model 1} and \textit{Model 3}, we follow \citet{2021JCAP...12..009M}, namely \(z\in[0.7,2.0]\) for the former and \(z\in[0.9,1.8]\) for the latter.

Operatively, we model the luminosity functions following the recipes outlined in \citet{2021JCAP...12..009M}. However, being the luminosity $L$ an intrinsic property of galaxies, we cannot measure it directly. Experimental observations usually deal with flux density or apparent magnitude, and it is hence useful to move from a luminosity-based to a flux-based description. To this purpose, we take into account the observed density flux $F$ as related to $L$ by the well-known inverse-square law \review{$F=L/(4\,\pi\,d_{\rm L}^2)$, where $d_{\rm L}(z)$ is the luminosity distance to redshift $z$}. For this reason, we use $F$, rather than $L$, to define galaxy samples.
Therefore, in our description we have a total sample---defined as the set of all the sources observed with a flux density $F\ge F_{\rm c}$---and the faint and bright subsamples---where $F\in[F_{\rm c},F_{\rm s})$ and $F \geq F_{\rm s}$, respectively.

Regarding the linear clustering bias, we adopt the phenomenological formula presented in \citet{2020MNRAS.493..747P}, which gives us the cumulative bias for $\rm H \alpha$ galaxies for both the total and the bright populations. (Recall that we can express the $b_{\rm F}$ as a function of  the other biases thanks to \cref{eq:Bias_faint}.)
Then, we follow the procedure outlined in \citealt{2021JCAP...12..009M} to obtain the magnification end evolution biases. As described above, we notice that the definitions of $\bmag_{\rm B}$ and $\bevo_{\rm B}$ are derived exactly as for the total sample by simply substituting $F_{\rm c}$ with $F_{\rm s}$. However, for the faint sample, the expressions are somewhat different, due to the presence of the upper cut \(F_{\rm s}\). Specifically, we write the following original relations for the faint sample,
\begin{align} 
    \bmag_{\rm F} &=\frac{y_{\rm c}\,g(y_{\rm s})-y_{\rm s}\,g(y_{\rm c})}{\displaystyle\int_{y_{\rm c}}^{y_{\rm s}}g(y)\,\de y}\;, \\ 
    \bevo_{\rm F} &=-\frac{\de\ln{\phi_*(z)}}{\de\ln{(1+z)}}-\frac{\de\ln{L_*(z)}}{\de\ln{(1+z)}}\,\bmag_{\rm F} \;.
\end{align}

\begin{figure}
\centering
\includegraphics[width=\columnwidth]{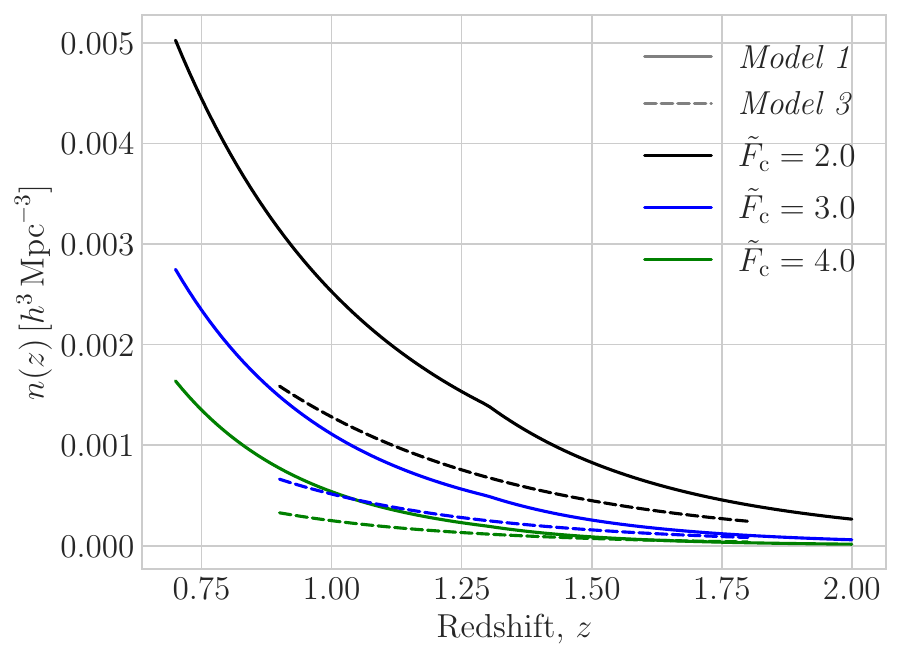}
\caption{Galaxy number density for a total/bright galaxy population (i.e.\ without the upper cut) in the case of the \textit{Model 1} (solid lines) and \textit{Model 3} (dashed lines) luminosity function models. Different colours represent different luminosity lower cuts, that is, black for $\tilde F_{\rm c}=2.0$, blue for $\tilde F_{\rm c}=3.0$ and green for $\tilde F_{\rm c}=4.0$---where $\tilde F=F/(10^{-16}\,\mathrm{erg\,cm^{-2}\,s^{-1}})$. As expected, the larger $F_{\rm c}$, the smaller $n(z;F>F_{\rm c})$.}
\label{fig:Euclid_n_g}
\end{figure}

\Cref{fig:Euclid_n_g} depicts the number density of sources for both \textit{Model 1} and \textit{Model 3}, with three minimum fluxes. All curves refer to a bright population---i.e.\ they do not undergo any upper cut in luminosity---as the presence of the faint sub-sample would have been trivial, being $n_{\rm F}$ the difference between $n_{\rm T}$ and $n_{\rm B}$. We note that \textit{Model 1} is more optimistic in estimating the number density with respect to \textit{Model 3}, for this reason \citet{2021JCAP...12..009M} use $F_{\rm c}=3.0\fluxum$ as a reference for \textit{Model 1} while they fix $F_{\rm c}=2.0\fluxum$ in \textit{Model 3} \citep{2024arXiv240513491E}.

\subsection{Bright galaxy survey}
The DESI BGS survey detects $r$-band magnitude bright sources up to $z \sim 0.5$. Since we have to know how to describe the number counts of sources and the linear galaxy bias depending on both redshift and magnitude limit, we coherently recover all those quantities thanks to the results shown in \citet{2023arXiv231208792S}. That paper provides us with a Halo Occupation Distribution (HOD) fit for BGS sample that takes into account the (absolute) magnitude of the sources. Starting from their main results, we are thus able to get the number density---whose logarithmic derivatives give us $\bmag_{\rm T,\, B}$ and $\bevo_{\rm T,\, B}$---as well as $b_{\rm T,\, B}$. In doing so, we switch to a $r$-band apparent magnitude-based formalism and consider a K correction, accounting for the redshifting effect on the band, which we model according to \citet{2021JCAP...04..055J}. We then redefine the magnification bias as
\begin{equation}
    \bmag_{\rm T} = \frac{5}{2}\, \frac{\partial \log_{10}n(z;m<m_{\rm c})}{\partial m_{\rm c}}\;,
\end{equation}
being $m$ the $r$-band magnitude and $m_{\rm c}$ the critical magnitude, in this description.
In our calculations, we parametrise the HOD smooth step function describing the occupation number of central galaxies as an error function \citep{2005ApJ...633..791Z} and do not use Eq.\ (2) of \citet{2023arXiv231208792S}.\footnote{ChangHoon Hahn and Christophe Yeche, private communication.}

Again, following \citet{2021JCAP...12..009M} we estimate the magnification and evolution bias for the total and the bright samples and then compute the values for the faint selection with \cref{eq:Q_F,eq:bevo_F}.

\begin{figure}
\centering
\includegraphics[width=\columnwidth]{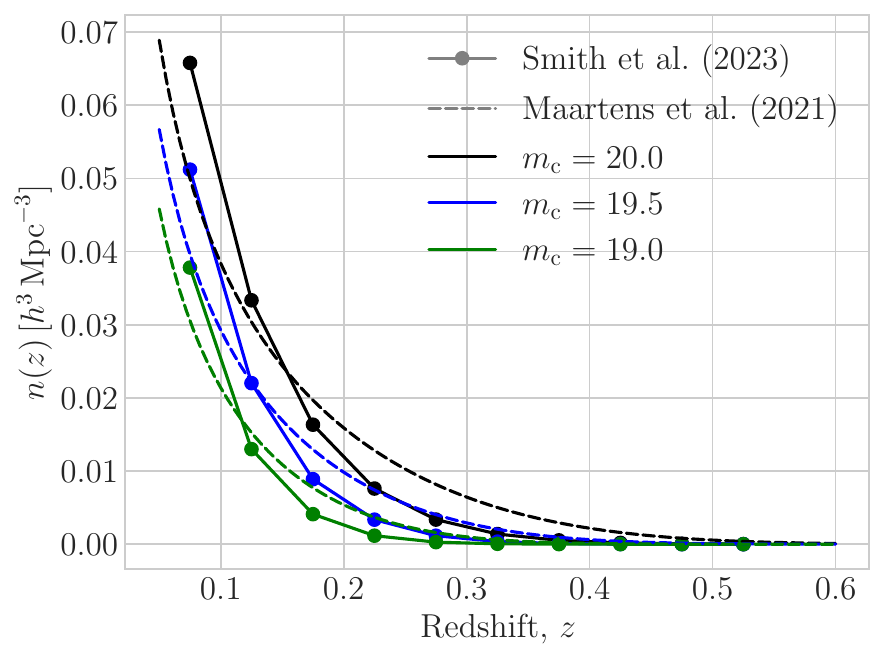}
\caption{Galaxy number density $n(z;m<m_{\rm c})$ for a DESI-like Bright Galaxy Sample. Solid lines refer to the number density obtained with our HOD-based approach---those used for the analysis---and dashed lines represent the analytical Schechter luminosity function, for comparison. As in \cref{fig:Euclid_n_g}, there is no upper luminosity cut---namely, no minimum apparent $r$-magnitude---and the colour code is: black for $m_{\rm c}=20$, blue for $m_{\rm c}=19.5$ and green for $m_{\rm c}=19$.}
\label{fig:BGS_n_g}
\end{figure}
Moreover, before computing any galaxy power spectrum, we assess the consistency of our HOD-driven approach by comparing its outcomes with a simpler analytical Schechter luminosity function model. We plot in \cref{fig:BGS_n_g} our $n_{\rm g}$---obtained following \citet{2023arXiv231208792S}, whose work relies on simulated data---as well as the number density given by the luminosity function in \citet{2021JCAP...12..009M}. The analytical model underestimates the number of objects at low redshift and, vice-versa, overestimates it at higher $z$ values. However, the agreement between the two approaches is enough to cross-validate the analytical description. We also point out that magnification and evolution bias plotted in \cref{fig:BGS_biases} are reasonably in line with Fig.\ 10 of \citet{2021JCAP...12..009M}.

\bibliographystyle{apsrev4-1}
\bibliography{apssamp}% Produces the bibliography via BibTeX.

\end{document}